\documentclass[conference]{IEEEtran}
\makeatletter
\newcommand\semihuge{\@setfontsize\semihuge{22.3}{22}}
\makeatother

\usepackage{algpseudocode}
\usepackage{algorithm}
\usepackage{algorithmicx}

\usepackage{lipsum} 
\usepackage{amsfonts}

\usepackage{multirow}
\usepackage[dvips]{color}
\usepackage{comment}
\usepackage{todonotes}
\usepackage{epsf}
\usepackage{epsfig}
\usepackage{times}
\usepackage{epsfig}
\usepackage{graphicx}
\usepackage{mathtools}
\usepackage{mathrsfs}
\usepackage{amssymb}
\usepackage{pdfpages}
\usepackage{epstopdf}
\usepackage{float}
\newfloat{algorithm}{t}{lop}
\usepackage{subfig}
\usepackage{dsfont}
\usepackage[multiple]{footmisc}
\usepackage{lettrine} 
\usepackage{amsmath,epsfig,amssymb,algorithm,algpseudocode,amsthm,cite,url}

\usepackage{caption}

\allowdisplaybreaks
\usepackage{csquotes}
\usepackage{multirow}

\usepackage{verbatim}
\usepackage{subfig}
\usepackage[english]{babel}
\usepackage{amsmath,amssymb}

\usepackage{verbatim}

\IEEEoverridecommandlockouts

\begin{document}
	\bstctlcite{IEEEexample:BSTcontrol}
	
	\title{\huge Convergence of Communications, Control, and Machine Learning for Secure and Autonomous Vehicle Navigation\vspace{-0.42cm}}
	
	\author{\IEEEauthorblockN{Tengchan Zeng\IEEEauthorrefmark{1}, Aidin Ferdowsi\IEEEauthorrefmark{1}, Omid Semiari\IEEEauthorrefmark{2}, Walid Saad\IEEEauthorrefmark{1}, and  Choong Seon Hong\IEEEauthorrefmark{3}}
	\IEEEauthorblockA{\IEEEauthorrefmark{1}\small{Wireless@VT, Department of Electrical and Computer Engineering, Virginia Tech, Arlington, VA, USA}}

		\IEEEauthorblockA{\IEEEauthorrefmark{2}\small{Department of Electrical and Computer Engineering, University of Colorado, Colorado Springs, CO, USA}} 
		
		\IEEEauthorblockA{\IEEEauthorrefmark{3}\small{Department of Computer Science and Engineering,
			Kyung Hee University, Yongin, South Korea}} 	
		E-mails: \IEEEauthorrefmark{1}\{tengchan, aidin, walids\}@vt.edu, \IEEEauthorrefmark{2}osemiari@uccs.edu,  \IEEEauthorrefmark{3}cshong@khu.ac.kr
		\vspace{-0.65cm}
	}
	
	\maketitle
	
	\begin{abstract}
		Connected and autonomous vehicles (CAVs) can reduce human errors in traffic accidents, increase road efficiency, and execute various tasks ranging from delivery to smart city surveillance.
		Reaping these benefits requires CAVs to autonomously navigate to target destinations. 
		To this end, each CAV's navigation controller must leverage the information collected by sensors and wireless systems for decision-making on longitudinal and lateral movements. 
		However, enabling autonomous navigation for CAVs requires a convergent integration of communication, control, and learning systems.
		The goal of this article is to explicitly expose the challenges related to this convergence and propose solutions to address them in two major use cases: Uncoordinated and coordinated CAVs.
		In particular, challenges related to the navigation of uncoordinated CAVs include stable path tracking, robust control against  cyber-physical attacks, and adaptive navigation controller design. 
		Meanwhile, when multiple CAVs coordinate their movements during navigation, fundamental problems such as stable formation, fast collaborative learning, and distributed intrusion detection are analyzed. 
		For both cases, solutions using the convergence of communication theory, control theory, and machine learning are proposed to enable effective and secure CAV navigation.
		Preliminary simulation results are provided to show the merits of proposed solutions.    
	\end{abstract}
	
	\IEEEpeerreviewmaketitle
	\section{Introduction}
\begin{figure}[!t]
	\centering 
	\includegraphics[width=2.5in,height=2.7in]{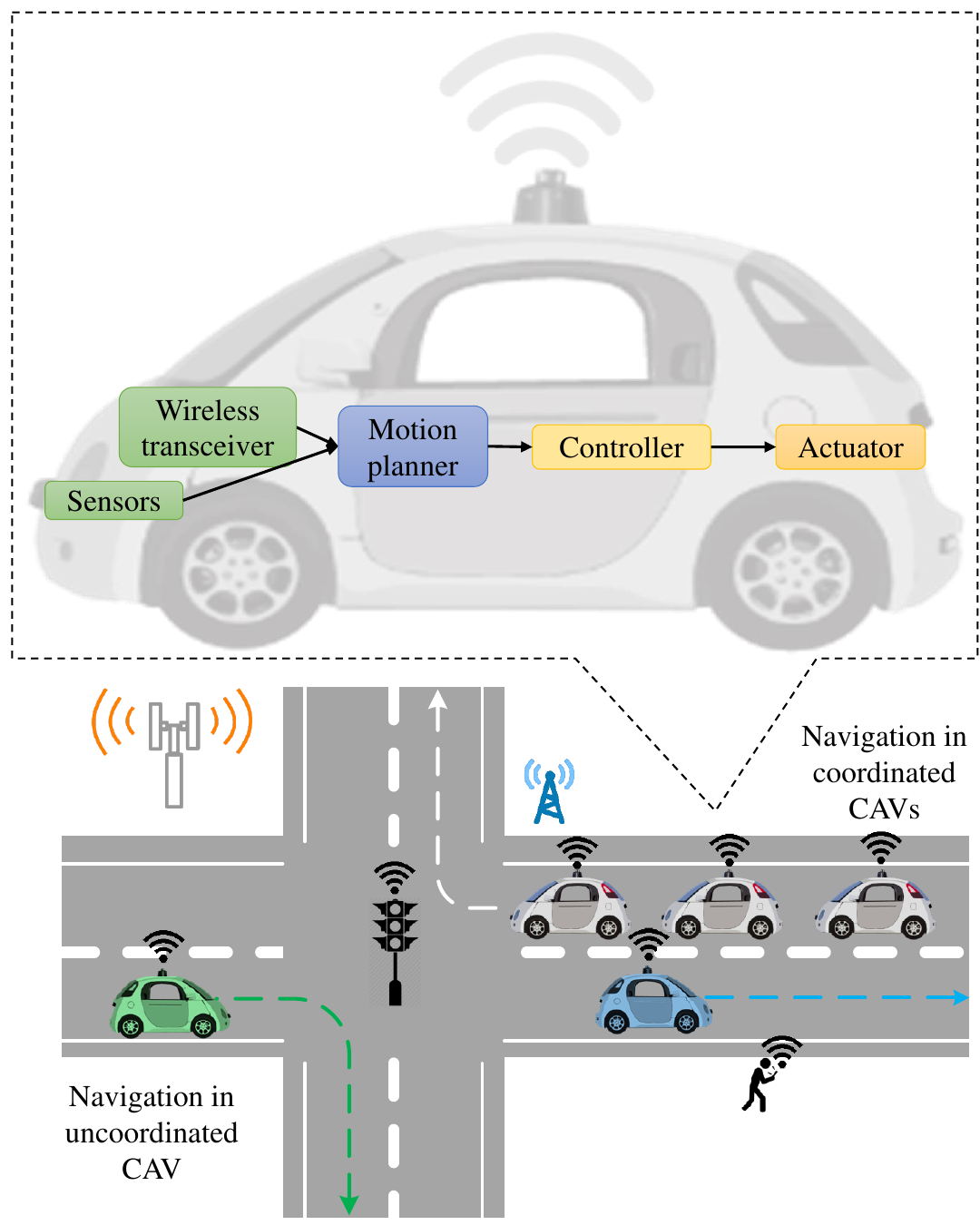}
	\vspace{-0.155cm}
	\DeclareGraphicsExtensions.
	\caption{Basic structure of a CAV and two CAV navigation use cases.}
	\vspace{-0.5cm}
	\label{basicStructure}
\end{figure}

	Connected and autonomous vehicles (CAVs) are promising solutions to reduce accidents, improve traffic efficiency, and provide various services, such as  delivery of medication using aerial CAVs. \cite{8457076}.
	To operate effectively, CAVs must perceive and sense their surrounding environment and autonomously navigate along a predesigned path to target destinations. 
 	As shown in Fig. \ref{basicStructure}, for a given CAV, environmental perception is accomplished by sensors and wireless connections with surrounding objects.   
	Along with the prior knowledge of the road network, such collected situational information will be processed by a motion planner to design target path and dynamics for navigation.  
	Subsequently, the controller will use the difference between the current
	dynamic parameters (e.g., location and heading angle) and desired targets designed at the motion planner as a feedback signal to make appropriate adjustments for the actuator commands.
	Such commands are later executed by the actuator that enables a CAV to track the target path and move to the destination.
	Depending on whether or not a CAV coordinates its movement with surrounding CAVs in the path-tracking process, autonomous navigation can be further studied in two use cases: Uncoordinated and coordinated CAVs. 

	However, unique system design and security challenges must be addressed for autonomous CAV navigation.
	For example, the CAV's wireless system is tightly integrated with its control and autonomy mechanisms.
	In particular, as shown in Fig. \ref{basicStructure}, the controller may heavily rely on the information collected by the wireless system. 
	Also, the communication network can transmit learning task data and learning models when CAVs use machine learning (ML) to complete navigation-related tasks like obstacle recognition. 
	With such integration, the controller operation and learning process of the CAV will be affected by the performance of the wireless system.
	\emph{Hence, for effective CAV navigation, we must determine how the wireless system affects the navigation controller and navigation-related learning tasks while jointly designing the integrated systems.}
Moreover, due to their reliance on both wireless networks and sensors (see Fig. \ref{basicStructure}), \emph{CAVs must be robust against cyber-physical attacks.}
Example attacks include malicious information injections and sensor data manipulations that help adversaries take control of the CAV. 
Because of the difference of autonomous navigation between uncoordinated and coordinated CAVs, the aforementioned challenges vary from one use case to another. 
Therefore, there is a need to study the interconnection between communications, control, and ML and use such interconnection to 
develop a convergent integration among interconnected systems for a secure and autonomous vehicle navigation system among uncoordinated and coordinated CAVs.

Although the wireless system is closely integrated with control and learning systems in CAV navigation, remarkably, most prior art studies each system separately. 
For example, the authors in \cite{damaj2022future} discuss future trends in the communications and processing technologies to enable CAVs, such as 5G and B5G requirements. 
In \cite{liu2023systematic}, a comprehensive and thorough overview of
the current state of vehicle control technology is presented with a focus on trajectory tracking control at the microscopic level and collaborative control at the macroscopic level.
The authors in \cite{8884164} surveyed different cloud-based and edge-based learning strategies that can assist the perception, mapping, and location for CAV navigation. 
Moreover, although many works looked at CAV path tracking security (e.g., see \cite{sun2021survey}, \cite{ju2022survey}, and references therein), these works often ignore the interdependence of cyber and physical systems in CAVs. 
Clearly, none of the prior works in \cite{damaj2022future,liu2023systematic,8884164,sun2021survey,ju2022survey} explicitly studied the close interconnection between systems for CAV navigation, as they often solely study one system and assume other systems to be blackboxes.

The main contribution of this article is a comprehensive study on the convergence of communications, control, and ML for secure and autonomous navigation in uncoordinated and coordinated CAVs. 
In particular, 
we analyze fundamental problems such as stable path tracking, robust control against cyber-physical attacks, and adaptive controller design for the navigation of uncoordinated CAVs. 
Meanwhile, for coordinated CAVs such as platoons and drone swarms, we investigate the key challenges of stable formations, fast collaborative learning, and distributed intrusion detection. 
We also provide preliminary results to showcase the benefits of the proposed solutions.
Finally, we present future research directions and open problems to further improve the joint system design. 
Note that although some recent surveys like \cite{10004548} looked at CAVs, they neither explicitly studied navigation in both uncoordinated and coordinated CAVs nor considered joint system design and cyber-physical attacks.

\section{Convergence of Communication, Control, and Learning for Uncoordinated CAVs}
\label{2}
We study uncoordinated scenarios where each CAV operates independently and tracks its own path without coordination with others. 
In particular, we investigate three key problems: stable path tracking, security, and adaptive navigation.

\begin{figure*}[t]
	\centering
\includegraphics[width=7in,height=2.9in]{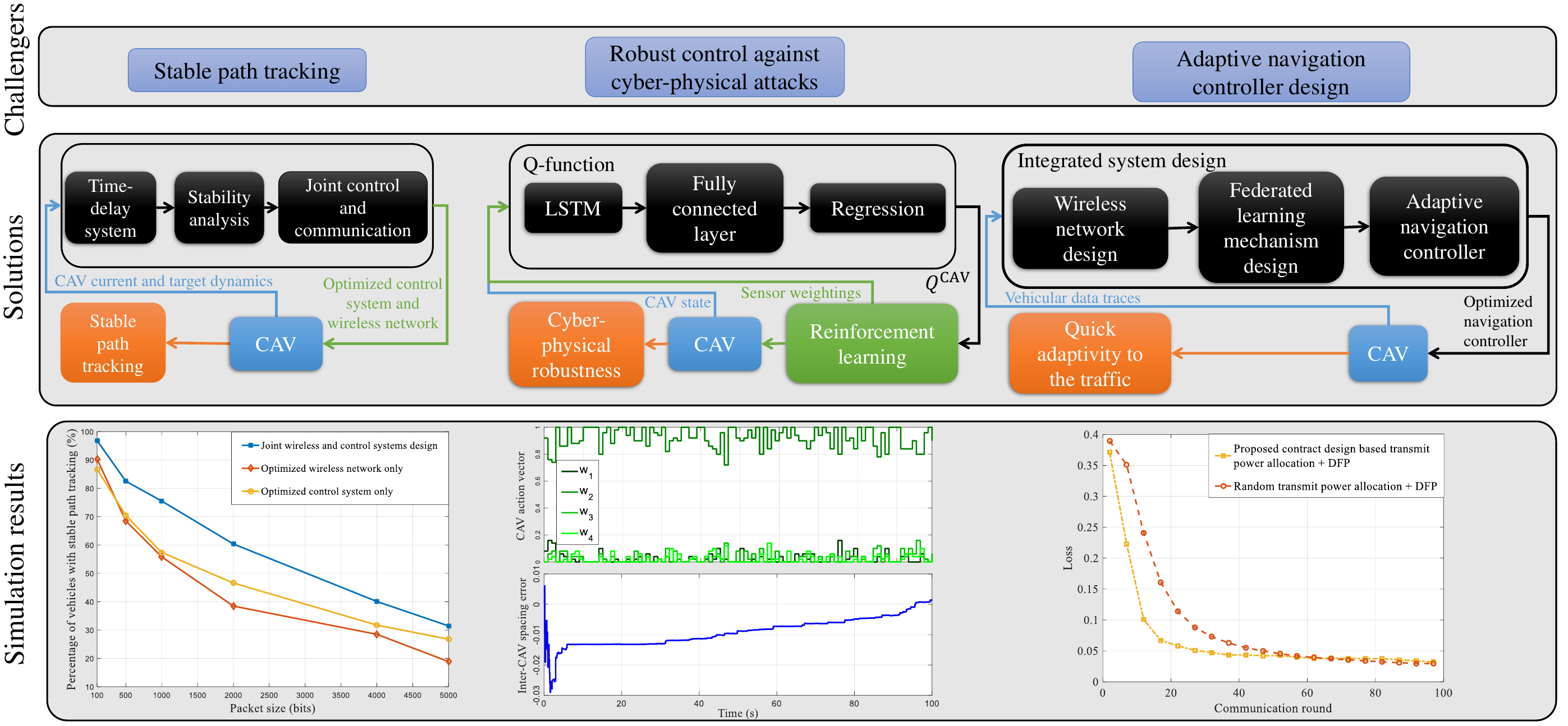}
\vspace{-0.155cm}
\DeclareGraphicsExtensions.
\caption{Challenges and their solutions to effective and secure navigation for uncoordinated CAVs.}
\vspace{-0.7cm}
\label{big1}
\end{figure*}
\subsection{Joint Communication and Control for Stable Path Tracking} \vspace{-0.02in}
\label{subsection2-A}
Although the environmental perception collected by wireless networks is critical for CAVs to decide the proper longitudinal and lateral movements for path tracking, such dynamic information will be inevitably subject to wireless delay and packet loss.
Here, delayed or wrong information can impair the stability of CAVs which is defined as the ability of CAVs to stabilize their dynamic parameters around targets. 
To better leverage wireless links for autonomous navigation and
path tracking, there is a need to determine the interconnections between the controller and the wireless network and, then, use them for joint communication and control design to ensure stable path tracking.

As a first step of joint system design, a time-delay system can be built to identify the synergies between control system and wireless network.
This is needed because a time-delay system can capture how dynamic errors between actual state and target state designed by motion planner change over time under the impact of wireless factors, like delay.
The \emph{Lyapunov-Krasovskii} and \emph{Lyapunov-Razumikhin theorems} \cite{gu2003stability} can then be used to build a stability analysis framework for the time-delay system. 
These theorems can help analyze the stability of time-delay systems and determine the wireless network requirements, such as delay threshold, which can prevent the instability of the
control system.
Finally, the wireless network and control system can be jointly optimized for stable path tracking. 
For example, the controller can choose appropriate values for control parameters in its control law to ease wireless network requirements.  
Meanwhile, the wireless network can optimize its design at the physical, link, and network layers so as to meet the control system's wireless network requirements.

The left block diagram in Fig. \ref{big1} shows how to perform a joint design of the communication and control systems for a stable path tracking. 
To test the effectiveness of this joint communication and control for path tracking, we simulate a Manhattan mobility model and a CAV that uses a pure pursuit controller for path tracking \cite{8761966}.
By using the block diagram on the left side of Fig. \ref{big1}, we derive the delay requirement imposed by the controller's stability.
For the joint system design, we use a dual method to derive the optimal headway distance for the pure pursuit controller. 
For the wireless system design, we leverage the conditional value-at-risk and branch and bound algorithm for allocating transmit power to maximize the number of vehicular communication links that meet the delay requirements.
As observed from the simulation results, the system with the joint design outperforms baselines optimizing the communication and control system separately. 
Clearly, we can see that the convergent integration between communication and control systems can yield a more reliable wireless network to support the stable navigation.

\subsection{ML for Robust CAV Control under Cyber-Physical Attacks}\vspace{-0.02in}
The reliance on sensors and communication links exposes a CAV to cyber-physical attacks by adversaries that seek to take control of the CAVs by remotely manipulating their data. Thus, the CAVs' data processing units must be robust to such attacks. One key attack is data injection attacks in which an adversary manipulates CAV sensor reading such that the CAV deviates from its designated path or spacing from its surrounding objects. This threat model is important to study because it can show the direct impact of a cyber attack on the physical operation of the CAV. Furthermore, as a robustness mechanism at the state estimation process, a CAV's data fusion center can assign different weights for its sensor. 
The state can include the CAV's location, speed, acceleration, and wheel angle. Therefore, the CAV's estimated state depends on the data fusion's weighting strategy and on the attacker's dynamic data injection strategy. To find a robust data fusion strategy, a natural way is to study the interaction between the attacker and the data fusion center using game theory.
However, in a CAV scenario, due to large and time-dependent CAV state space, game-theoretic analysis is challenging \cite{Ferdowsi2018}.

Instead, ML tools such as long-short term memory (LSTM) cells and reinforcement learning (RL) can be used to derive an effective CAV data fusion strategy \cite{Ferdowsi2018}. An LSTM block is a deep recurrent neural network that can store information for long periods of time and, thus, can learn long-term dependencies within a sequence.  Therefore, LSTMs are useful in learning temporal interdependencies between the CAV's state values from previous time steps and summarizing the the CAV's motion. Then, RL can be used by the fusion center to decide on the best action to choose based on the LSTM summary. The RL component seeks to find a data fusion strategy that minimizes the long-term effect of the data injection attack. Training this combined LSTM and RL scheme allows designing a data fusion center for CAVs that can robustly control the CAV and estimate its state.

The middle block diagram in Fig. \ref{big1} shows an LSTM-based robust CAV controller from \cite{Ferdowsi2018}. In the simulation, we study a data injection attack on a CAV within a scenario in which each CAV has four sensors and the attacker can inject data to sensors 1, 3, and 4 to cause deviation in the spacing between CAVs on a street. We observe that the LSTM based controller can detect the attack on sensors thus setting the weight of sensor 2, $w_2$, close to 1 while the other weights will be set to values close to 0. 
Therefore, by using LSTM to extract temporal features from CAVs' state and RL to determine data fusion strategy, one can identify the cyber-physical interdependence in attacker's actions and goals and develop a convergent approach to ensure that CAVs' motion becomes robust against security attacks.
\subsection{FL for Adaptive Navigation Controller Design}
Another key CAV navigation challenge is designing controllers that can accurately execute real-time control decisions. 
Here, conventional feedback controllers can fail to adapt the CAVs to various road types, dynamic road traffic, and varying weather and payloads, since they are usually designed for a fixed vehicle model and road condition. 
Meanwhile, even if popular learning methods (e.g., neural networks) can be used to design adaptive navigation controller, the local data can be insufficient and skewed to train the learning model. The reasons are due to the limited on-chip memory available on board CAVs and the fact that an individual CAV can only store data pertaining to its most recent travels. 
Therefore, a distributed ML framework among multiple CAVs will be needed for properly designing CAVs' adaptive navigation controller.

To this end, one can leverage the wireless connectivity in CAVs and use \emph{federated learning (FL)} to enable a group of CAVs to jointly train the learning models used by their controllers. 
In FL, these CAVs can train the controller models based on their local data and, then, a parameter server, e.g., a base station (BS), can aggregate the trained controller models from CAVs. 
This process will be repeated among the CAVs in consecutive rounds and parameter server iteratively until all controllers converge to the optimal learning model \cite{konevcny2016federated}. As such, \emph{the learning model can be trained among multiple CAVs, and such a trained model can enable a particular CAV’s controller to adapt to new traffic scenarios previously unknown to it but experienced by other CAVs.}
To guarantee a good convergence, the interconnection between the wireless, learning, and control systems must be considered. 
Specifically, the wireless network must be designed to achieve a high participation of CAVs in FL even in the presense of mobility and channel uncertainty.
Meanwhile, since the data quality among CAVs change from one CAV to another, the FL algorithm design must account for unbalanced and non-independent and identifcally distributed (non-IID) local data across CAVs. Here, the CAVs can effectively converge to using the optimal navigation controller and quickly adapt to the dynamic road traffic.

The right block diagram in Fig. \ref{big1} shows an autonomous controller design for CAVs. 
In the simulation, we assume that the CAV uses an adaptive proportional-integral-derivative (PID) controller to adjust longitudinal movement where the control parameters are tuned by an artificial neural network (ANN) auto-tuning unit. 
To train the ANN model parameters, we design a novel FL algorithm, i.e., dynamic federated proximal (DFP) algorithm, that accounts for CAV mobility, fading, and the unbalanced and non-IID data across CAVs. 
Moreover, an incentive mechanism is designed to determine the transmit power allocation strategy and increase CAVs' participation in FL. 
We use real vehicular data traces (i.e., Berkeley deep drive \cite{fyu2020}) for our framework and we can observe that the convergence
speed of the DFP algorithm with our incentive mechanism design can be improved by 40\% compared with baseline, justifying the merits of considering the convergent integration between
the wireless network, FL, and control system for adaptive navigation controller design.

\section{Convergence of Communication, Control, and Learning for Coordinated CAVs}
\label{3}
Next, we discuss autonomous navigation for coordinated CAVs where a group of CAVs coordinate their movements to complete sophisticated missions in an uncertain environment.  
Example applications are CAV platoons and drone swarms.
In a CAV platoon, to increase the road capacity, multiple CAVs operate together and maintain small spacing between each other.  
In a drone swarm, a group of aerial CAVs fly along predesigned paths while keeping a target formation to perform collaborative tasks, e.g., cooperative obstacle recognition. 
In such applications, several joint system design and security challenges must be addressed: stable formation, fast collaborative learning,
and distributed intrusion detection.
\begin{figure*}[t]
	\centering
		\includegraphics[width=7in,height=2.9in]{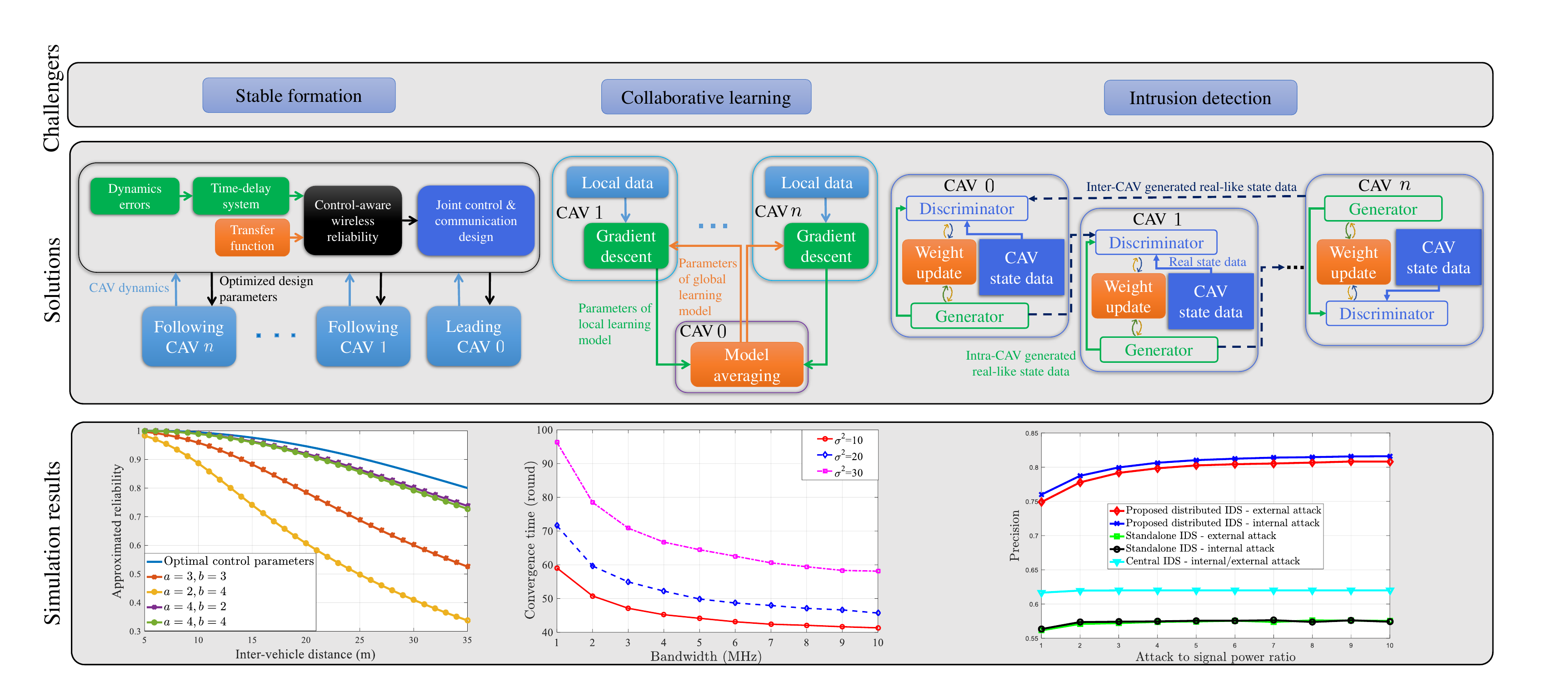}
		\vspace{-0.155cm}
		\DeclareGraphicsExtensions.
		\caption{Challenges and their solutions to effective and secure navigation for coordinated CAVs.}
		\vspace{-0.7cm}
		\label{big2}
\end{figure*}

\subsection{Joint Communication and Control for Stable Formation}
\label{stableFormation}
For operational safety,  autonomous navigation usually requires  coordinated CAVs to maintain a stable formation. 
For example, a stable formation for a platoon means that each CAV maintains a target speed and safe spacing to its preceding CAV. 
To achieve such formations, CAVs will use sensors and wireless systems to collect dynamics (e.g., speed and location) of nearby CAVs.  
As already discussed, the wireless delay and packet loss can jeopardize the CAVs' stable operation.
Hence, there is a need to study how the wireless network affects the stable formation for coordinated CAVs.  

Similar to Section II-A, a time-delay system can be built for the group of CAVs so that the delayed dynamic errors (such as velocity and spacing errors) are explicitly considered. 
To analyze the impact of the wireless network on the stable formation, there are two types of stability to consider for coordinated CAVs: Plant and string stability. 
To secure plant stability where all CAVs maintain the same speed and a target distance from surrounding CAVs, the Lyapunov-Krasovskii and Lyapunov-Razumikhin theorems can be used.
Meanwhile, in order to guarantee string stability where the disturbances (e.g., in
velocity or distance) of nearby vehicles are not amplified
in the CAV group, transfer function from control theory can be used. 
This is due to the fact that the transfer function can capture how the disturbances propagate among adjacent coordinated CAVs. 
In this case, a non-increasing disturbance
can thereby equal to a monotonically decreasing transfer
function in coordinated CAVs.
Based on plant and string stability analysis, the delay threshold can be derived so that coordinated CAVs can operate with a stable controller.
This characterized delay can enable the joint communication and control system design where the design of the control system is optimized to relax the delay
constraints on the wireless network and the wireless network can be optimized to improve its reliability to support controller's stability.  

The left block diagram in Fig. \ref{big2} shows the joint communication and control design which keeps the stable formation for a group of $n+1$ coordinated CAVs.
In the simulation, we consider a highway scenario for a platoon where the acceleration of each platoon CAV is determined by a proportional controller and the distribution of vehicles generating interference to platoon CAVs is modeled by stochastic geometry \cite{8778746}. 
In particular, we calculate the control delay requirement and  
the wireless reliability whose performance is shown in the simulation under different proportional controller parameters $a, b \!\in\! [2,4]$ \cite{8778746}.
We can observe that the platoon with the optimized controller design can achieve a higher wireless reliability for wireless networks compared with platoons with other control parameters.
Clearly, the controller design directly affects how reliable the wireless networks is to support a stable formation of coordinated CAVs. 
In other words, a stable formation of coordinated CAVs will require solutions using the convergence of
communication theory and control theory.
\subsection{Joint System Design for Fast Collaborative Learning} 
\label{III-A}
ML will play an important role to execute collaborative navigation-related tasks such as cooperative obstacle recognition for coordinated CAVs (note that CAVs must be coordinated to keep safe spacing and ensure operational safety while learning). 
One common approach is using cloud-based ML where the CAV data is sent to a cloud for data analysis and inference. 
However, continuously transferring the data of many CAVs to cloud servers will not be resource-efficient if even feasible, since it will impose an unmanageable traffic load over the access network.
Also, as the shared data (like trajectory information) can be privacy-sensitive, malicious servers can intrude the CAVs' privacy, e.g., by tracking the CAV's location. 

To facilitate collaborative ML while maintaining the privacy of CAV navigation data, FL can be used.
Essentially, FL divides coordinated CAVs into a global CAV and multiple local CAVs and uses an iterative update scheme whose details can be found in Section II-C. 
However, as FL models are transmitted via communication links, the learning performance hinges on underlying wireless network \cite{TLi}.
For example, due to the limited battery life and transmit power, not all CAVs can participate in the FL
training and some important local model updates can be discarded, leading to a poor execution performance for these navigation-related tasks. 
Meanwhile, the network design must take into account the communication requirement imposed from the control system to maintain operation safety. 
Hence, the impact of wireless network on the convergence of FL and safe spacing between CAVs must be explicitly considered when using FL for collaborative navigation related tasks. 

The middle block diagram in Fig. \ref{big2} shows the FL implementation over $n+1$ coordinated CAVs. 
Here, we implement FL for a swarm of wireless-connected drones with the leader as the global CAV and followers as local CAVs. 
Then, we perform a rigorous convergence analysis to derive the convergence time while considering wireless factors, e.g., random antenna angle deviations $\sigma^2 $, and the minimum number of CAVs to achieve a target convergence.
We observe from the simulation results that, when the variance of antenna angle deviations increases, FL needs more time to converge. 
This is because, when the variance increases, the antennas at the transmitting and receiving drones will be less aligned, reducing the antenna gain and increasing the transmission delay.
These results show that, when CAVs use FL for collaborative navigation, the FL performance directly depends on the underlying wireless setup.  
This connection is the foundation for convergently integrated system design when optimizing FL performance for coordinated CAVs.

\subsection{Distributed Generative Adversarial Networks for Intrusion Detection in Multi-agent CAVs}
The operation of coordinated CAVs requires addressing security challenges with a minimum delay to maintain coordination. By manipulating a CAV's controller in coordinated CAVs,
an adversary can control other CAVs as well thus causing cascading failures. To detect intrusions in multi-agent CAVs such as CAV platoons or drone swarms, an intrusion detection system (IDS) is typically implemented at a central node such as a BS that receives CAV state information. However, such centralized IDSs cannot detect intrusion or anomalies on-time due to the large-scale nature of multi-agent CAVs and communication delays. Therefore, in coordinated CAVs, IDSs must operate in a distributed fashion with minimum dependence on a central node. 

Generative adversarial networks (GANs) are ANN architectures that can be used as IDSs. A GAN architecture consists of 1) a \emph{generator} ANN that tries to generate real-like anomalous data samples and 2) a \emph{discriminator} ANN that discriminates between the anomalous data samples generated by the generator and the real normal data samples. The generator trains the discriminator against anomalies by generating numerous anomalous data samples which helps the discriminator explore the anomaly space. This GAN architecture can be implemented in a distributed fashion in multi-agent CAVs where every CAV owns a discriminator and a generator. Next, each CAV can train its ANN by its normal state information and share its ANN parameters with other CAVs using FL. Here, each CAV learns its own normal state and the other CAVs's normal state. Using this framework each CAV will own a discriminator that monitors its own and neighboring CAVs' behavior in coordinated tasks without a central node.

The right block diagram in Fig. \ref{big2} shows a distributed GAN-based IDS for CAVs. Specifically, we consider that each CAV uses a GAN to learn its own and neighbors' normal state and uses this GAN to detect anomalies and intrusions. We simulate an internal attack (attack on a given CAV state) and an external attack (attack on the state of the neighboring CAVs of a given CAV) \cite{FerdowsiGAN}.  We observe that a distributed GAN-based IDS has a higher precision of detecting internal and external attacks on CAVs compared to central and standalone IDSs. 
This is because a distributed IDS can identify interconnection between cyber and physical systems among CAVs and allow each CAV to learn normal state from not only itself but also other neighboring CAVs.
However, standalone IDSs do not consider neighboring CAVs and centralized IDS considers all of the CAVs as a system and neglects unique behaviors of each CAV. Therefore, by using GANs, we can implement distributed IDSs for coordinated CAVs such that each CAV can detect intrusions to the neighboring CAVs without any dependence on a central units which reduces communication delays.\vspace{-0.05in}

\section{Future Directions and Open Problems on Joint System Design and Security Design}\vspace{-0.02in}

\subsection{Joint System Design for Advanced Controller Design}\vspace{-0.03in}
Classical control laws are usually used to deal with nominal operating conditions, like flat roads for ground vehicles and large and open spaces for
aerial vehicles. However, to deal with more extreme operating conditions, like urban areas, there is a need for more sophisticated models for the controller design, like rear wheel position based feedback. When designing these advanced controllers, it is gaining
momentum to use the machine learning framework for training its local data and optimizing the
controller model. In this case, all the challenges pertaining to the interconnection between communication, control and learning will still persist, i.e., the impact of wireless network on learning performance and controller' stability, as well as a proper learning mechanism design for CAVs to facilitate a fast convergence to the optimal controller model. 
Hence, the goal of this future work is to use the interconnection between control, learning, and wireless 
framework for the advanced controller design.\vspace{-0.05in}
\subsection{Advanced FL for CAVs}\vspace{-0.05in}
As an extension of FL used in Sections \uppercase\expandafter{\romannumeral2}-C and \uppercase\expandafter{\romannumeral3}-B, advanced FL strategies can be further used to optimize the CAV system design.
For example, hierarchical FL can be used to aggregates the knowledge learned by more CAVs.
In particular, each BS will update the model, called intermediate model, based on model updates received from its associated CAVs, and BSs will also send the updated intermediate model to a central entity, hosted on an edge or a regional cloud server, who will generate a new
global model and transmit the new global model to BSs.
Also, since conventional FL frameworks suffer from straggler effects and partial participation of CAVs in the learning process, there is an increasing interest on asynchronous FL where the BS will aggregate the received parameters and generate a new global model in an asynchronous fashion. 
Same to the conventional FL used in Sections II-C and III-B, the impact of wireless network on the convergence performance must be explicitly studied for advanced FL frameworks.
Also, when used for the navigation controller design, these advanced FL mechanisms must be properly designed to consider the unbalanced and non-IID data, the CAVs' mobility, and the wireless fading channels.\vspace{-0.05in}
\subsection{Adversarial ML in CAVs}\vspace{-0.05in}

As shown in Section \ref{III-A}, CAVs can rely on ML to complete tasks like object detection, path planning, and trajectory optimization. Such reliance on ML make CAVs vulnerable against adversarial ML that attempt to alter ML models through malicious inputs. For example, adversarial ML techniques can fool a CAV such that it will not detect an object on its path which would cause an accident. This is because the ML models used in CAVs are usually trained on stationary and benign environments such that the behavior of the environments is assumed to be unchanged after the ML models are trained. However, the presence of an intelligent adversary may disturb the normal behavior of the CAV environment such that the pretrained ML model would fail to fulfill its task. Techniques such as adversarial training or defensive distillation are two of the commonly used methods against adversarial ML. Adversarial training uses a lot of adversarial examples and explicitly train the model not to be fooled by each of them. Moreover, in defensive distillation, the output of an ML model will be probabilistic rather than hard decisions which will make it difficult for the adversary to find adversarial inputs that lead to incorrect decisions by CAVs. Therefore, adversarial ML for CAVs is a crucial research area for CAV decision  making and operation.
\vspace{-0.05in}

\section{Conclusions}\vspace{-0.02in}
In this article, we have highlighted the joint system design and security challenges in two use cases of coordinated and uncoordinated CAV navigation.
Using the convergence among communication theory, control theory, and ML, we have proposed solutions to address key challenges, such as stable path tracking, robust control against cyber-physical attacks, and navigation controller design for uncoordinated CAVs, as well as challenges pertaining to coordinated CAVs, including stable formation, fast collaborative learning, and distributed intrusion detection systems. Simulation results have been provided to verify the merits of proposed solutions.   
Finally, we have outlined some of key open problems further optimizing the joint system design and addressing security challenges for the CAV navigation.
The convergence study in this article can be extended to any interconnected systems within CAVs and can be used to optimize the CAVs' operations, beyond navigation.
\def\baselinestretch{0.7}
\bibliographystyle{IEEEtran}

\begin{IEEEbiography}[{\includegraphics[width=1in,height=1.25in,clip,keepaspectratio]{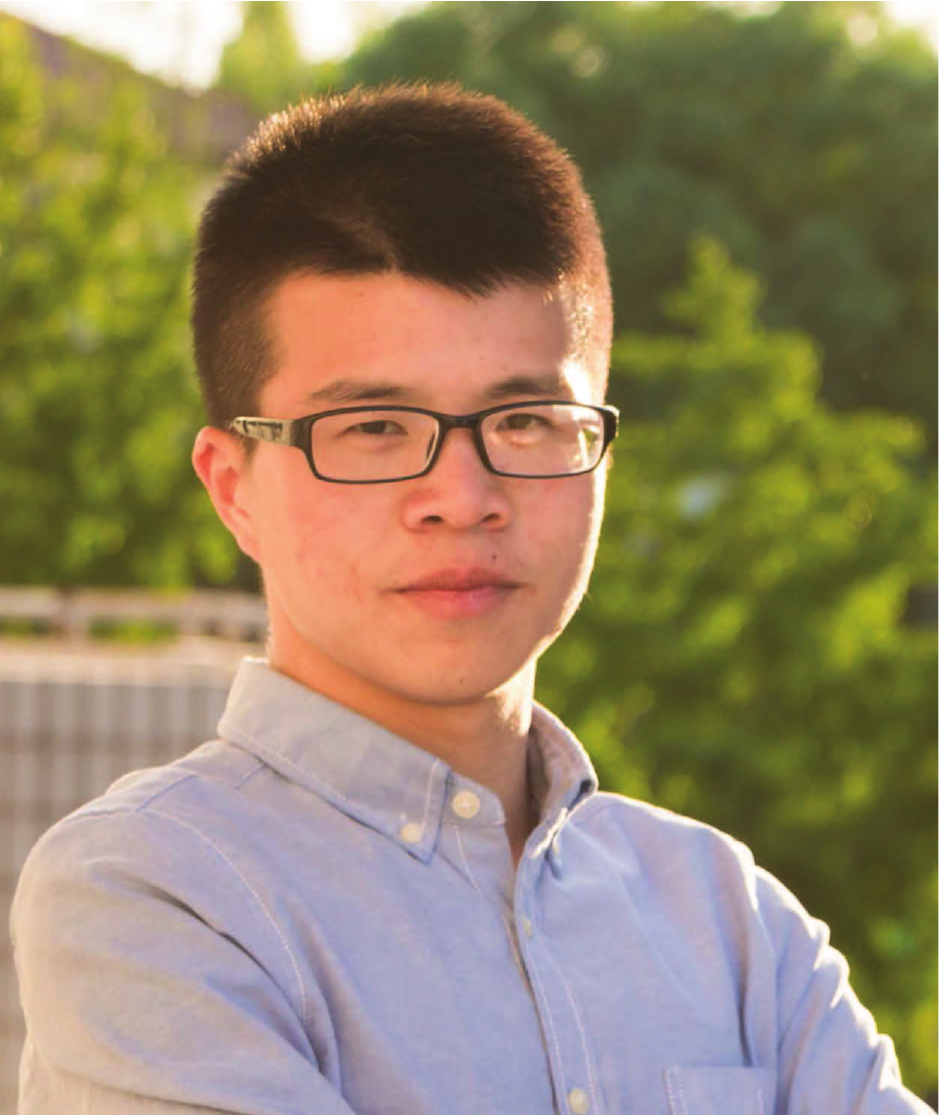}}] {Tengchan Zeng} (S'18) received his Ph.D. degree from Virginia Tech in 2021. He is currently a Systems Engineer at Ford Motor Company. He was an exemplary reviewer for IEEE Transactions on Communications in 2021.  \vspace{-1cm}
\end{IEEEbiography}

\begin{IEEEbiography}[{\includegraphics[width=1in,height=1.25in,clip,keepaspectratio]{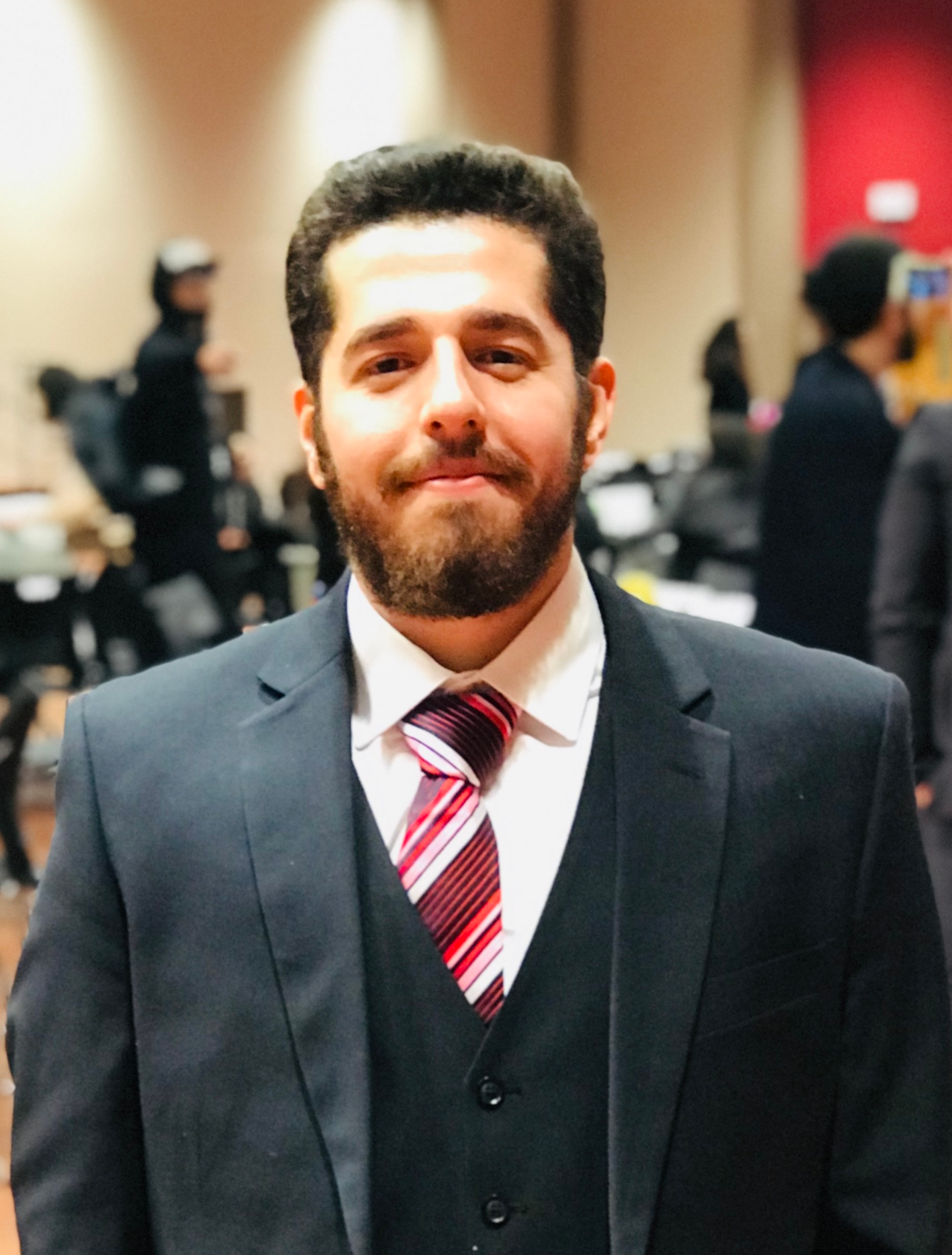}}]{Aidin Ferdowsi} (S'17) received his Ph.D. degree in electrical engineering from Virginia Tech. He is currently a Principal Machine Learning Engineer at Capital One. Dr. Ferdowsi is awarded The Outstanding Dissertation Award in all STEM majors from Virginia Tech. His research interests include machine learning, data science, and game theory. \vspace{-1cm}
\end{IEEEbiography}

\begin{IEEEbiography}[{\includegraphics[width=1in,height=1.25in,clip,keepaspectratio]{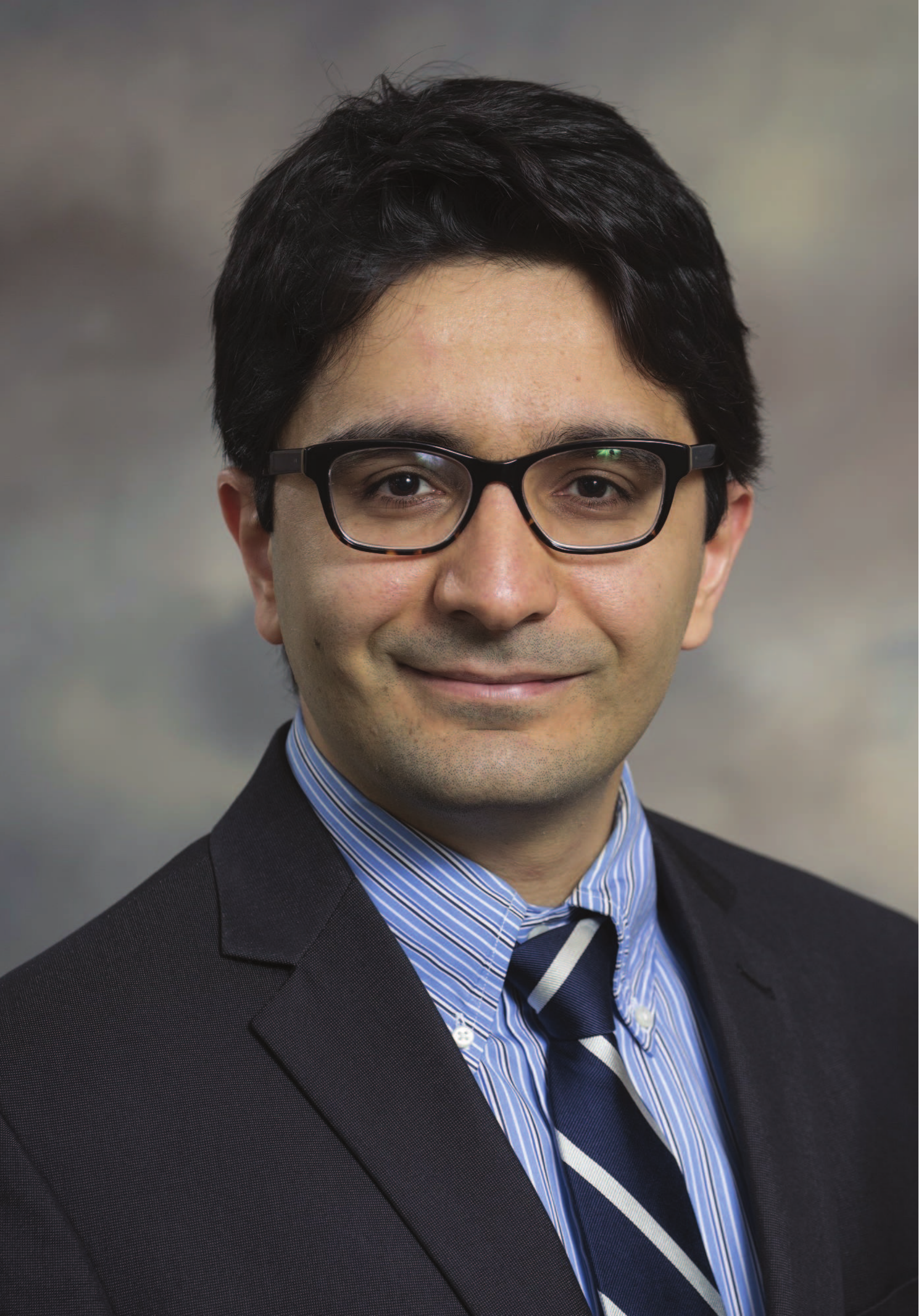}}] {Omid Semiari} (S'14, M'18) received his Ph.D. from Virginia Tech in 2017. He is an assistant professor with the ECE department at the University of Colorado, Colorado Springs. His research interests include wireless communications (6G), machine learning for communications, distributed learning, and cross-layer network optimization. \vspace{-1cm}
\end{IEEEbiography}

\begin{IEEEbiography}[{\includegraphics[width=1in,height=1.25in,clip,keepaspectratio]{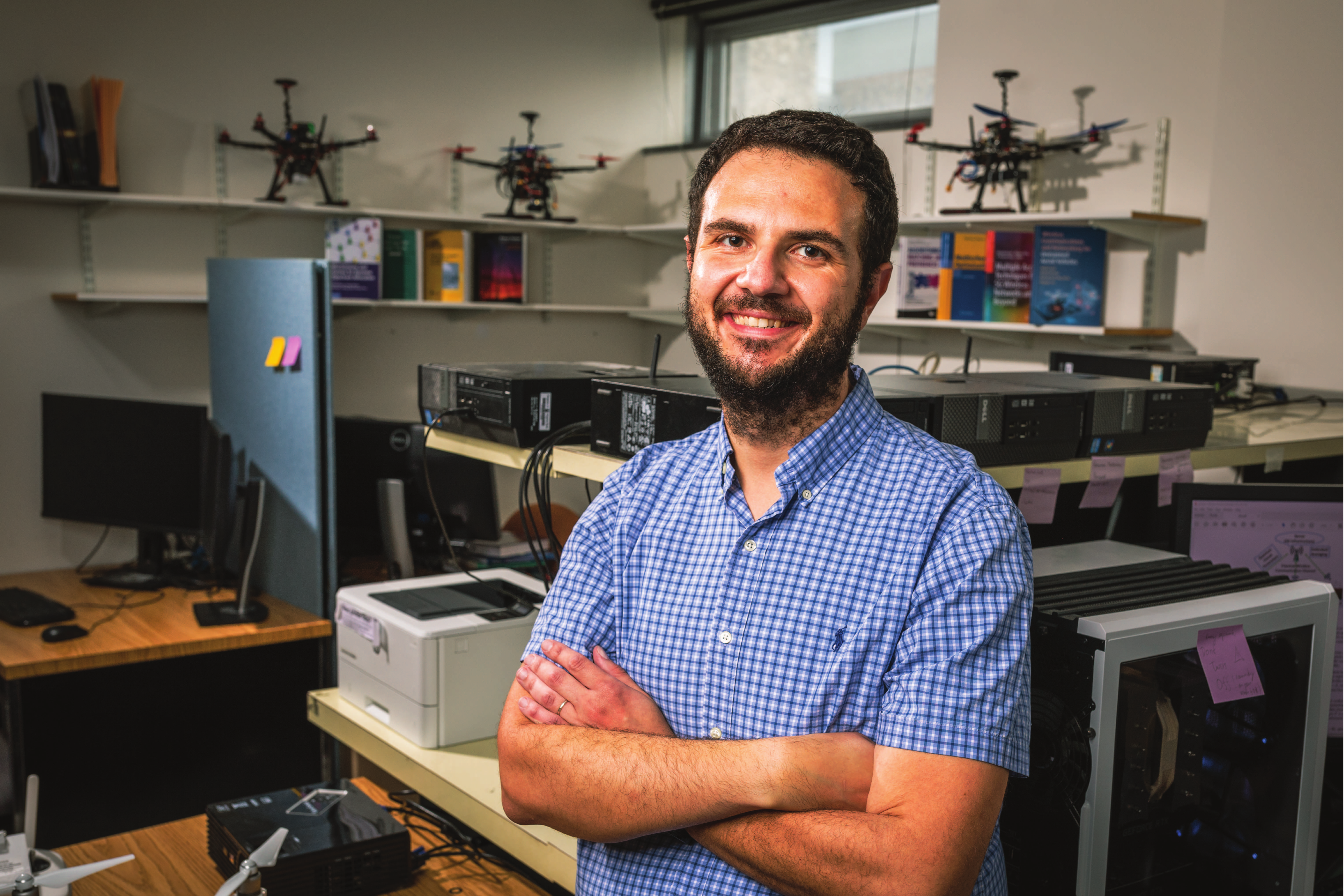}}] {Walid Saad} (S'07, M'10, SM'15, F'19) received his Ph.D degree from the University of Oslo in 2010. Currently,  he is a Professor at the Department of Electrical and Computer Engineering at Virginia Tech where he leads the Network sciEnce, Wireless, and Security (NEWS) laboratory. His  research interests include wireless networks, machine learning, game theory, cybersecurity, unmanned aerial vehicles, cellular networks, and cyber-physical systems. Dr. Saad was the author/co-author of eleven conference best paper awards  and of the 2015 IEEE ComSoc Fred W. Ellersick Prize. He is a Fellow of the IEEE.  \vspace{-1cm}
\end{IEEEbiography}

\begin{IEEEbiography}[{\includegraphics[width=1in,height=1.25in,clip,keepaspectratio]{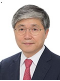}}]{Choong Seon Hong}  (S'95, M'97, SM'11) is working as a professor with the Department of Computer Science and Engineering, Kyung Hee University. His research interests include machine learning, mobile computing, federated learning and satellite networking. He was an Associate Editor of the IEEE Transactions on Network and Service Management, J. Communications and Networks and an Associate Technical Editor of the IEEE Communications Magazine. \vspace{-1cm}
\end{IEEEbiography}

%
%
%
\end{document}